\documentclass[preprint,times]{aastex}
\usepackage{natbib}

\usepackage{graphicx}
\usepackage{epsf}

\begin{document}

\title{Near Infrared Observations of Comet-Like Asteroid (596) Scheila}
\author{Bin Yang$^{1}$ and Henry Hsieh$^{1,2}$ 
\affil{$^1$Institute for Astronomy, University of Hawaii, 2680 Woodlawn Drive, Honolulu, HI
96822 \\
$^2$Hubble Fellow \\} 
\email{yangbin@ifa.hawaii.edu, hsieh@ifa.hawaii.edu}}

\begin{abstract}
Asteroid (596) Scheila was reported to exhibit a cometary appearance and an increase in brightness on UT 2010 December 10.4. We used the IRCS spectrograph on the 8-m Subaru telescope to obtain medium-resolution spectra of Scheila in the HK-band (1.4 - 2.5$\mu$m) and low-resolution spectra in the KL-band (2.0 - 4.0$\mu$m) on UT 2010 December 13 and 14. In addition, we obtained low-resolution spectroscopy using the SpeX spectrograph on the 3-m NASA Infrared Telescope Facility (IRTF)  telescope on UT 2011 January 04 and 05.  The spectrum of Scheila shows a consistent red slope from 0.8 to 4.0$\mu$m with no apparent absorption features, resembling spectra of D-type asteroids. An intimate mixing model suggests that the amount of water ice that might be present on the surface of Scheila is no more than a few percent. The spectrum of the Tagish Lake chondrite matches the asteroid's spectrum at shorter wavelengths ($\lambda < 2.5 \mu$m), but no hydration features are observed at longer wavelengths on Scheila. Our analysis corroborates other studies suggesting that the comet-like activity of Scheila is likely not caused by the sublimation of water ice. The dust coma and tail may be results of a recent impact event. 

\end{abstract}

\section{Introduction}
Asteroid (596) Scheila was reported to show a cometary-like coma and an increase in brightness on UT 2010 December 10.4 \citep{larson:2010}. The reported sudden activity of (596) Scheila shares several similarities with cometary outbursts, which often involve the development of dust and ice comae and a sudden increase in brightness. Scheila, with a semi-major axis of 2.93 AU, is located in the outer asteroid main belt, where several main-belt comets \citep[MBCs;][]{hsieh:2006} have been found. Observations of recurrent activity in two MBCs indicate that the activity is likely driven by the sublimation of water ice \citep{hsieh:2004, hsieh:2010, hsieh:2011a}. Recently published thermal models \citep{schorghofer:2008, prialnik:2009} show that subsurface ice at heliocentric distances similar to those of the MBCs could survive billions of years if it is protected from direct sunlight by refractory dust mantles $\sim$ 100-meters thick. To date, no water ice has ever been directly detected on MBCs \citep{rousselot:2011}. However, space missions (Deep Space I, Stardust, Deep Impact, EPOXI) show that nuclei of active comets are largely covered by porous, dark, refractory materials \citep{Keller:2004} and only a trace of water ice was detected on the surface of comet 9P/Tempel 1 \citep{sunshine:2007}. Detecting water ice on a main belt object from the ground is therefore clearly a challenging task, and as such, the lack of detections to date is unsurprising. 

Available observations of the MBCs are mainly in the optical. 133P/Elst-Pizarro is the only object that has been studied in the near-infrared \citep[NIR;][]{rousselot:2011}. Because of the small sizes of MBCs (sub-kilometer to a few kilometers in diameter), most MBCs are simply too faint for NIR observations and even fainter at longer wavelengths ($\lambda$ $>$ 2.5$\mu$m). Pioneering studies of asteroids have recognized that observations in the 3$\mu$m region provide the best chance of detecting diagnostic features \citep{lebofsky:1980, jones:1990,rivkin:2002}. The 3$\mu$m region is particularly important because a combination of bending and stretching modes of the O-H bond in water as well as hydroxyl-bearing minerals produce several strong features in this region. Unlike absorption features at shorter wavelengths, 3$\mu$m bands are often saturated and detectable even at small ($\sim$2 wt\%) concentrations of absorbing materials \citep{jones:1990}. Recently, a broad absorption band centered at 3.1$\mu$m was reported on asteroid (24) Themis \citep{rivkin:2010,campins:2010}, the parent body of the Themis family with which two known MBCs are associated \citep{hsieh:2006}. The band center and the shape of the absorption feature has been reported to be consistent with the diagnostic feature of water ice. The detection of water ice on the largest family member (24) Themis, if real, would strongly support the hypothesis that MBCs could have been produced through the fragmentation of an ice-rich parent body. However, water ice is not the only material that can explain the 3$\mu$m feature. The spectrum of Goethite (FeO(OH)), an iron oxide, matches the 3$\mu$m feature in the spectrum of Themis adequately well \citep{Beck:2011}. 

Based on the Eight-Color Asteroid Survey \citep[ECAS;][]{zellner:1985}, Scheila was originally classified as P- or D-class asteroid \citep{tholen:1989}. More recently, it was classified as a T-type asteroid based on the Small Main-Belt Asteroid Spectroscopic Survey II \citep[SMASS II;][]{bus:2002}, which has a higher spectral resolution than the ECAS survey. T-type and D-type asteroids, though classified into two distinct groups, are very similar in that both appear red and featureless in the optical. The only difference between these two types lies in the wavelength region from 0.85 to 0.92$\mu$m, where the T-types show flatter spectra than the D-types \citep{bus:2002}. We note that the red spectral slope and the low geometric albedo, $p_{v}$=0.038 $\pm$ 0.004 \citep{tedesco:2002}, of Scheila closely resembles  that of several comet nuclei \citep{AHearn:1988, Licandro:2003, Abell:2005}. If its surface properties are indicative of intrinsic composition, water ice may be present in Scheila. If the observed cometary activity is powered by the sublimation of water ice, then icy grains may be found in the vicinity of the nucleus. For example, water ice features were detected in the spectra of outbursting comets 17P/Holmes and P/2010 H2 Vales, which were observed a few days after their initial outbursts \citep{yang:2009, yang:2010}.  Thus, we promptly performed NIR spectroscopy of Scheila to search for diagnostic absorption features of water ice, on the surface and vicinity of Scheila. In this paper, we present the results of that search. 

\section{Infrared Observations}
We obtained NIR spectroscopy of Scheila on UT 2010 December 13 and 14 using the 8.2-m Subaru telescope.  On UT 2011 January 04 and 05, we made additional observations using the 3.0-m NASA Infrared Telescope Facility (IRTF) atop Mauna Kea. During our Subaru observations, a 188-element curvature sensor adaptive optics system (AO188) was used \citep{minowa:2010}. Given Scheila's brightness (V $\sim$ 13), we were able to operate the AO188 system in the natural guide star (NGS) mode, using the asteroid itself as a guide ``star''. With the AO correction, the seeing was improved to $< 0\farcs1$ in the K-band. We observed Scheila in two spectroscopic modes. First, we adopted the 52 mas/pix mode and the HK-grism, which provides a spectral coverage from 1.4 to 2.5$\mu$m and a spectral resolution of $R \sim600$. Second and more importantly, we observed Scheila in the L-band (2.8 - 4.0$\mu$m). We adopted a low spectral resolution KL prism and the 20 mas/pix mode that provides a spectral coverage from 2.0 to 4.0$\mu$m and a spectral resolution of  $R \sim$ 250. The advantage of using the KL prism is that we are able to observe the target in the K-and the L-band simultaneously, greatly minimizing the possibility of false detections due to mis-alignments of the two spectral parts if they had been obtained separately. No coma or tail structures were observed in the J- and K-band images. We therefore adopted the A-B dither pattern, which is to nod the telescope along the slit by 3$''$. The slit width was fixed to $0\farcs23$ and a north-south slit alignment was used for all of our observations. Nearby G-stars were observed both for approximate removal of telluric absorption features and for removal of the solar spectral profile. We observed both Schila and G-type stars (HD 91163, HD 76332) near their transits, which occurred at similar airmasses. 

IRTF observations were made using a medium-resolution 0.8-5.5$\mu$m spectrograph (SpeX) \citep{rayner:2003}. We adopted the high throughput prism mode (0.8 - 2.5$\mu$m) and a $0\farcs8$x15$^{\prime\prime}$ slit that provides a spectral resolution of $R \sim$ 100, and kept the slit oriented along the parallactic angle to minimize effects from differential atmospheric refraction. G-type stars (HD 91162, HD 87680) were observed before and after the asteroid observations on each night.  A journal of observations is provided in Table 1. 

SpeX data were reduced using the SpeXtool reduction pipeline \citep{cushing:2004}, which follows standard data reduction procedures.  The Subaru data were processed following the standard spectroscopic data reduction procedures based on the Image Reduction and Analysis Facility (IRAF), which are described in the Subaru data-reduction cookbook on the IRCS website\footnotemark. 
 \footnotetext{http://subarutelescope.org/Observing/DataReduction/}

 \section{Results}
 \subsection{Absence of Absorption Features}
\begin{figure}[h]
\vspace{0.5 cm}
\hspace{-1 cm}\includegraphics[width=6in,angle=0]{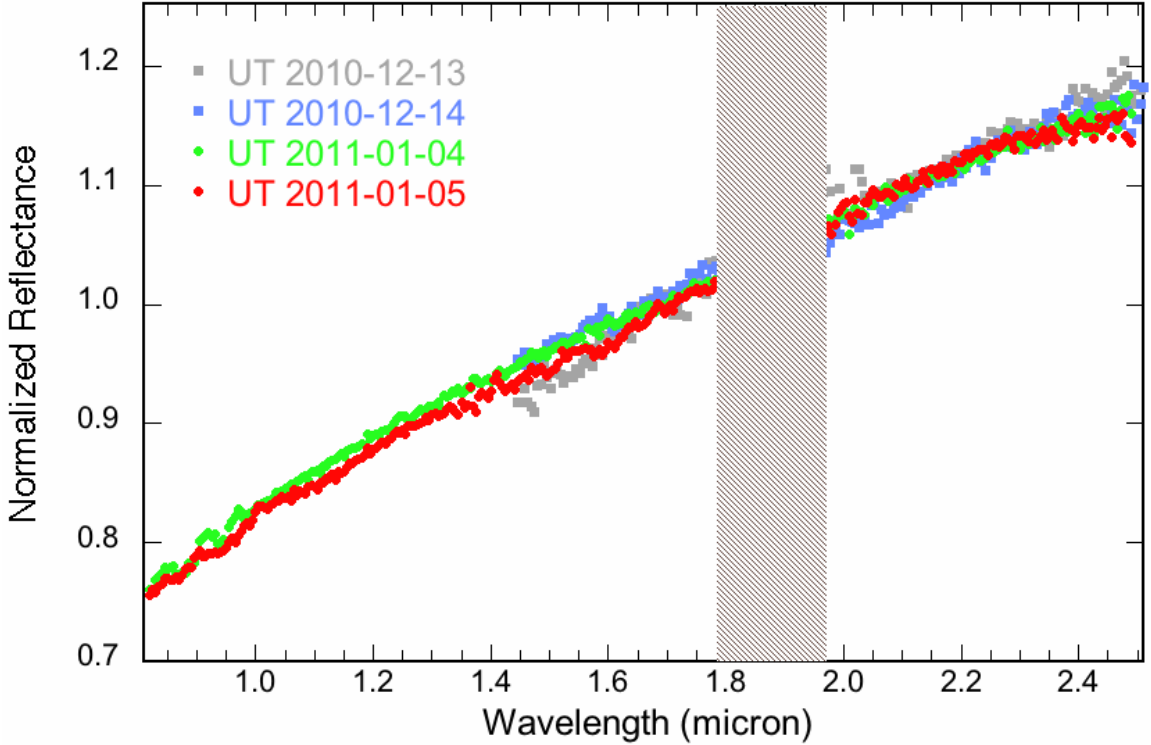}
\caption{NIR spectra of Scheila. Data taken with Subaru on UT 2010 December 13 and 14 are marked with grey and blue solid squares, respectively.  Solid green and red dots mark data taken with IRTF on UT 2011 January 04 and 05. The scatter of points from the continuum near 2.0 $\mu$m is caused by the residual of imperfect removal of telluric absorptions.}
\label{all_fig1}
\end{figure}

Our Subaru and IRTF observations of Scheila in the spectral range 0.8$\mu$m $< \lambda <$ 2.5$\mu$m are presented in Figure \ref{all_fig1}. Although the four spectra were taken at different times and with different instruments, all the spectra consistently show linear spectra with reddish slopes. The Subaru spectra (solid squares) were binned to a lower spectral resolution (R $\sim$ 100) that is equivalent to the resolution of  the IRTF spectra (solid dots). Narrow small features between 1.4$\mu$m and 1.8$\mu$m and below 1.0$\mu$m in the IRTF spectra and the scatter of points away from the continuum near 2.0$\mu$m in the Subaru data are due to imperfect removal of telluric absorptions. No diagnostic features are detected at the 2\% noise level. 

\begin{figure}[h]
\begin{center}
\vspace{-1.5 cm}
\includegraphics[width=6.in,angle=0]{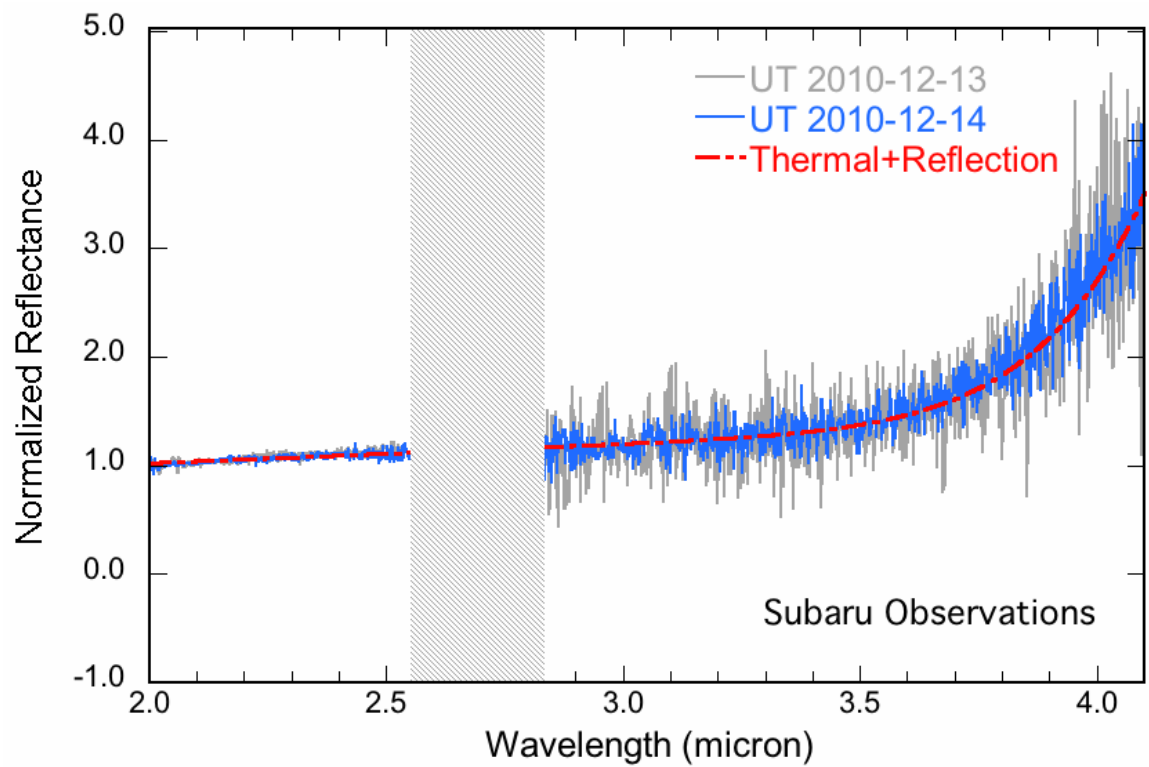}
\caption{Grey and blue solid lines mark spectra taken with Subaru in the KL-band on UT 2010 December 13 and 14. A simple thermal model, marked with a red dashed line, provides a good match with the thermal excess observed beyond 3.5$\mu$m.}
\label{KL_raw}
\end{center}
\end{figure}

The Subaru observations in the KL-band are shown in Figure \ref{KL_raw}. The two spectra taken on consecutive nights, shown in blue and grey, consistently show moderate reddish slopes from 2.0 to 3.5$\mu$m. Beyond 3.5$\mu$m, both spectra show a sharp rise which is caused by thermal emission from the surface of the asteroid. To remove the thermal excess, a simple thermal model is used. The reflection is modeled as a linear function of wavelengths and the thermal portion is calculated using a modified blackbody function:
\begin{equation}
f_{BB}(\lambda) = \frac{\epsilon \pi B_\lambda(T_e) R_{e}^{2}}{\Delta^2}
\end{equation}
\noindent where $R_{e}$ is the effective radius of the object and $B_{\lambda}(T_c)$ is the Planck function evaluated at the effective temperature and $\epsilon$ is the emissivity. Given Scheila's effective radius \citep[$R_{e}$  = 56.7~km;][]{tedesco:2002} and assuming $\epsilon=1$ \citep{morrison:1973, fernandez:2003}, we find a best-fit effective temperature of $190\pm10$~K for the thermal excess, shown as the red dashed line in Figure~\ref{KL_raw}. Water ice and hydroxyl-bearing minerals, if present, would produce broad absorption bands in the spectral region from 2.7 to 3.4$\mu$m \citep{rivkin:2002}, which is free from thermal contamination. As shown in Figure \ref{KL_raw}, the observed reflectance spectra show linear red slopes and no absorption features.

\subsection{Constraints on Surface Water Ice}
\begin{figure}[h]
\begin{center}
\includegraphics[width=4.5in, angle=90]{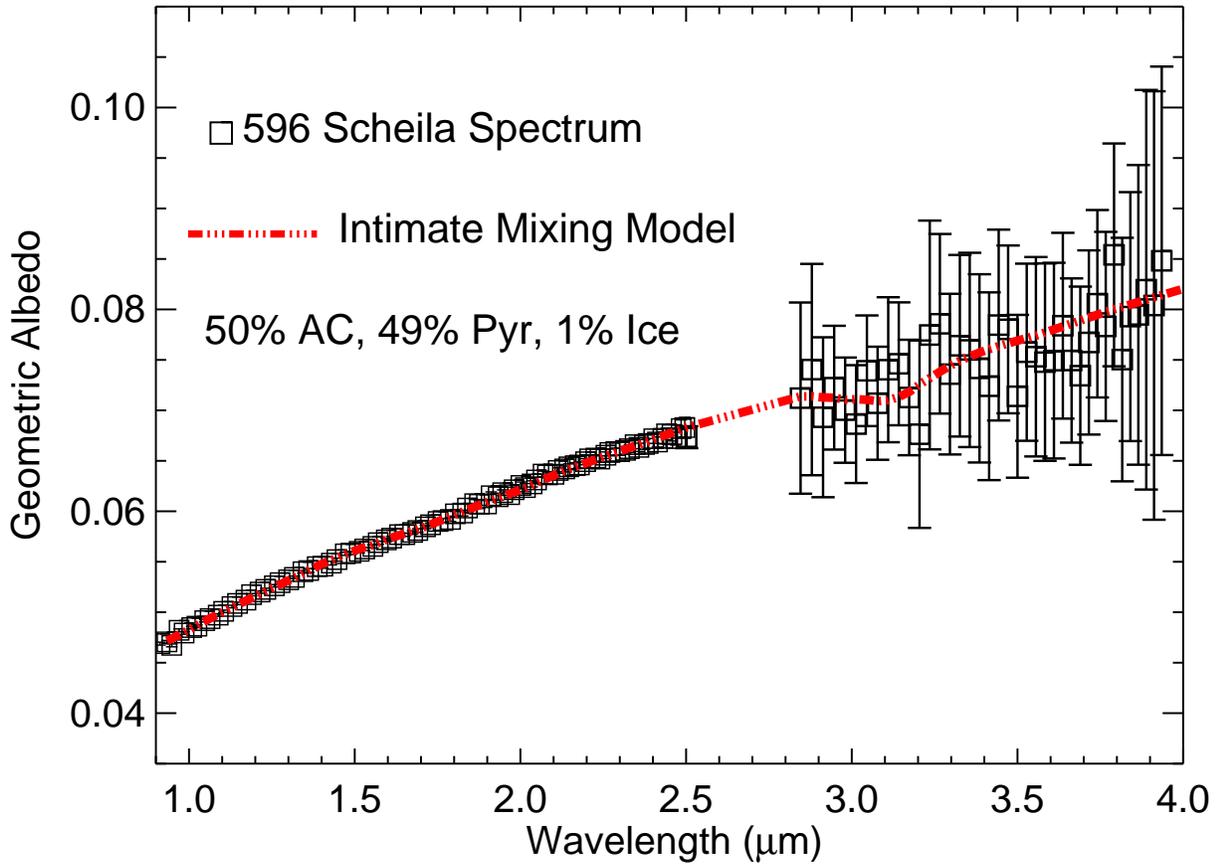}
\caption{Open squares are the Subaru observation of Scheila on UT 2010 December 14. The red dashed line is the best-fit Hapke radiative transfer model. AC stands for amorphous carbon and Pyr stands for pyroxene. No absorption features are observed. Our model suggests that surface water ice is no more than 1 wt\% on this object. }
\label{ice_model}
\end{center}

\end{figure}

To estimate an upper limit to the amount of water ice on Scheila's surface, we use an intimate mixing code to simulate the spectrum of the asteroid over 1.0 - 4.0$\mu$m. This code, developed and provided by Ted Roush, calculates synthetic geometric albedo spectra using the Hapke formalism \citep{hapke:1981,hapke:1993} and assuming isotropic scattering from all grains. We adopt three compositionally distinct components: water ice, amorphous carbon, and iron-rich pyroxene. We note that the choice of amorphous carbon and pyroxene is not unique. In principle, any combination of dark and red components that are spectrally featureless at 1.0 - 4.0$\mu$m could produce a similar fit when mixed with a small amount of water ice. In our models, the optical constants (OCs) of water ice are taken from \citet{warren:1984}, those for amorphous carbon are from the astronomical laboratory of the Astronomical Institute of the University of Jena \citep{Henning:1999}, and those for pyroxene are from \citet{dorschner:1995}. Our best-fit model (red dashed line in Figure \ref{ice_model}), consists of 50 wt\% amorphous carbon ($\bar{d} = $1.2$\mu$m), 49 wt\% pyroxene ($\bar{d} = $3.8$\mu$m) and 1 wt\% water ice ($\bar{d} = $2.6$\mu$m), and matches the overall shape of the asteroid spectrum (open squares) quite well. Although the mixing model only contains 1\% micron-sized water ice, a shallow absorption feature of water ice is clearly visible in the synthetic spectrum. This broad absorption band is not detected in our Subaru data. Given the signal-to-noise ratio (SNR) of the data, however, we can not rule out the presence of water ice at a level of a few percent. We note that the optical constants of water ice used in this study are measured at a temperature (266K) that is higher than the actual surface temperature ($\sim$190K) of Scheila. The temperature difference however does not affect the result of our spectral fitting. \cite{mastrapa:2009} investigated the effects of temperature on the absorption features of water ice. They found that temperature has the least effect on the  3.1 micron band. The band center only shifts about 0.01$\mu$m between 20K and 150K and the band depth changes by a few percent. Given the low spectral resolution and the limited SNR of our Subaru data, such small shifts would not have been discernible in our data. Our spectral models show that the surface of Scheila probably consists of fine-grained regolith with an average grain size of a few microns. 

\subsection{Comparison with Meteorite Analogs}
\begin{figure}[h]
\begin{center}
\hspace{-1 cm}
\includegraphics[width=3.1in,angle=0]{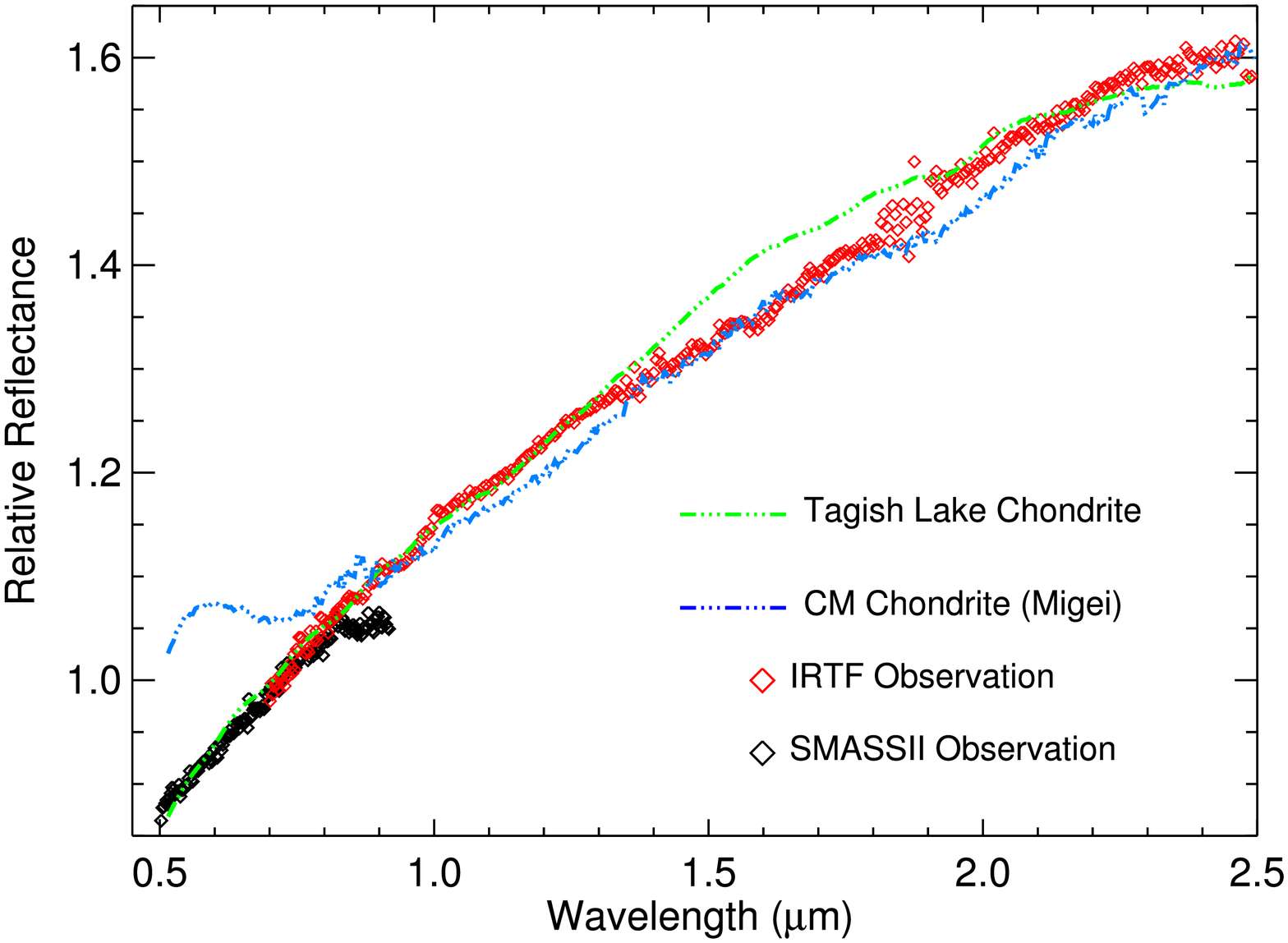}\hspace{-0.2 cm}\includegraphics[width=3.1in,angle=0]{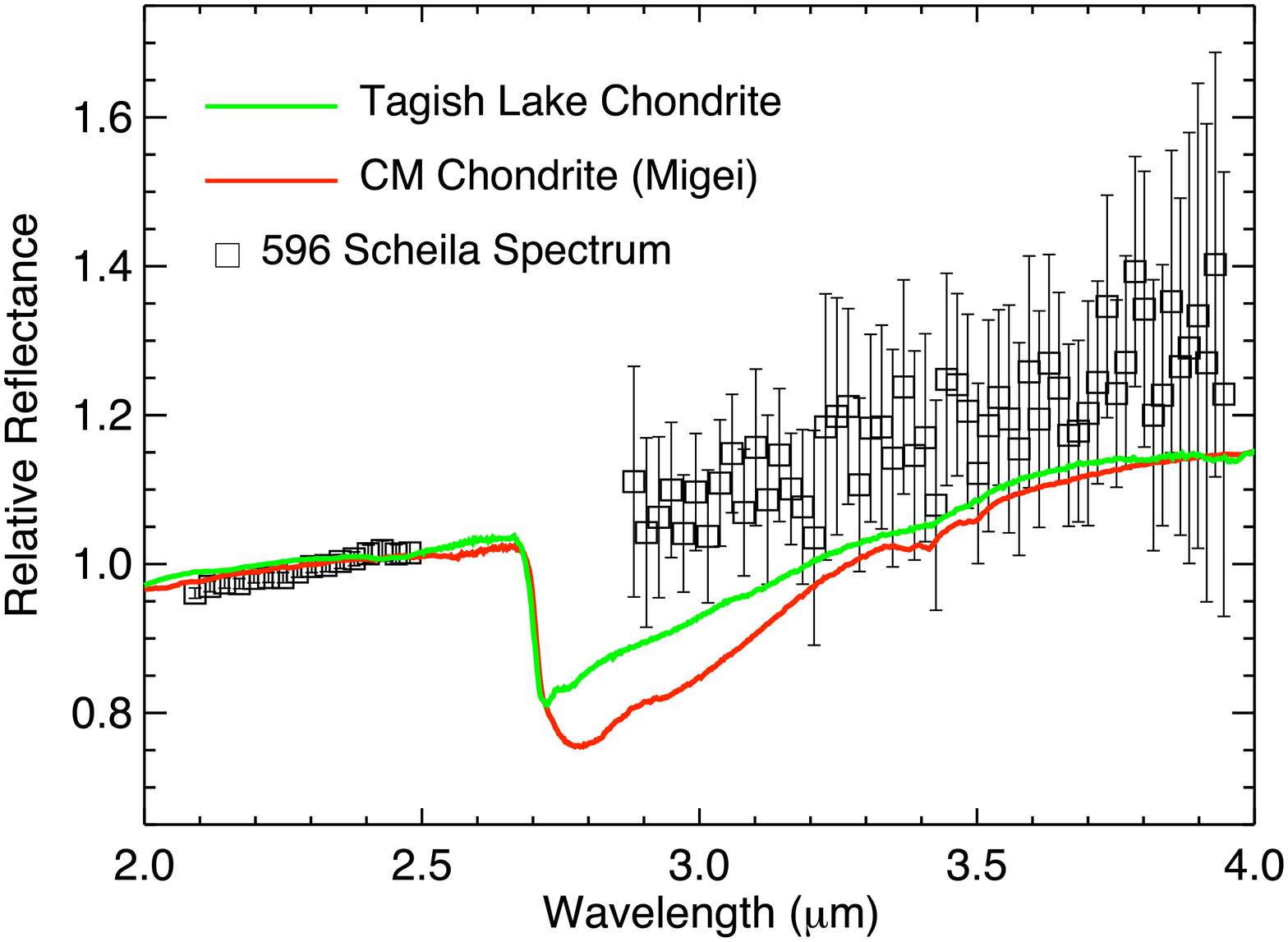}
\caption{$Left:$ Red open diamonds mark NIR observations of Scheila obtained in this study. Black open diamonds mark visible observations of Scheila taken from the SMASS II survey \cite{bus:2002}. The NIR spectrum and the visible spectrum are normalized and merged at 0.75$\mu$m. Green and blue dashed lines mark the spectra of carbonaceous chondrites. The spectrum of the Tagish Lake chondrite fits the asteroid spectrum better than a CM chondrite spectrum but a clear discrepancy is seen from 1.5 to 2.0$\mu$m. $Right:$ Spectra of Scheila (open squares) and chondrites (colored solid lines) in the KL-band. Both chondrites exhibit an absorption band centered at 2.7$\mu$m and 2.8$\mu$m, respectively. This hydration feature is not present in the asteroid spectrum.  }
\label{f4}
\end{center}
\end{figure}

Although no NIR absorption features were detected, the profiles of the spectra still hold valuable information about the composition of the Scheila's surface material. As such, we searched for spectral analogs for Scheila among meteorite samples in the Relab spectral library.  First, we combined the optical spectrum of Scheila taken from the SMASS II survey with the NIR spectrum obtained in this study. As shown in Figure \ref{f4}a, the spectral slope of the newly obtained NIR spectra (red open diamonds) is a good match to the SMASS II visible spectrum (black open diamonds) from 0.7$\mu$m to 0.8$\mu$m. We note that our IRTF spectrum of Scheila appears linear and red from 0.85$\mu$m to 0.92$\mu$m, and the previously observed flattened turnover is not confirmed in our data. Our observations show that the red and featureless spectrum of Scheila is consistent with spectra of D-type asteroids. 

At wavelengths below 2.5$\mu$m, we found the spectrum of the Tagish Lake carbonaceous chondrite (green dashed line) to be the best fit to Scheila's spectrum. In addition, the albedo of Tagish Lake at 0.55$\mu$m is about 3\% \citep{hiroi:2003}, which is consistent with the visual geometric albedo of Scheila p$_v$ = 0.038$\pm$0.004 \citep{tedesco:2002}. Another chondrite, an irradiated CM chondrite Mighei (blue dashed line), is able to fit the asteroid spectrum from 1.0 to 2.5$\mu$m but significant discrepancy occurs at wavelengths less than 1.0$\mu$m. However, both carbonaceous chondrites are aqueously altered and show strong hydrous (OH) features near $2.7-2.8~\mu$m, and these features are completely absent from the spectrum of Scheila (Figure \ref{f4}b). At present, no meteorite analog is found that completely matches the spectrum of Scheila over the whole spectral range from 0.4 to 4.0$\mu$m.  In spite of the non-detection of ice features on the surface, the long-recognized spectral and albedo similarities between D-type asteroids and comet nuclei suggests that preserved ice could be present in Scheila's interior. Even if deeply buried ice exists, we do not believe that ice sublimation is the cause of Scheila's comet-like outburst as discussed in the following section.

\section{Discussion}
\subsection{Scheila and Known MBCs}
Although Scheila orbits in the main asteroid belt and exhibited temporary comet-like dust tails, it differs from other MBCs in several aspects. First, previous photometric and spectroscopic studies of MBCs have found that the optical and NIR colors of two nuclei of MBCs are comparable or slightly bluer than the Sun \citep{hsieh:2006, hsieh:2009b, jewitt:2009, licandro:2011, rousselot:2011}. For example, the reported spectral slope of 133P in the visible is S$^{\prime}_v$= 0.00 $\pm$ 0.01 \%/1000 \AA\ and that of 176P is S$^{\prime}_v$= -0.03 $\pm$ 0.01 \%/1000 \AA\ \citep{licandro:2011}. In contrast, our observations and the SMASS II observations of Scheila show that its spectral slope in the visible is S$^{\prime}_v$= 6.1 $\pm$ 1.0 \%/1000 \AA\, which is about 6 $\sigma$ higher than 133P and 176P.  Second, the equivalent circular diameter of Scheila is $D = 113 \pm2$~km \citep{tedesco:2002}, about 30 times the diameter of the largest MBC 176P \citep[$D=4.0\pm0.2$~km,][]{hsieh:2009a}.  According to \cite{bottke:2005}, D $>$ 110 km asteroids in the main belt are expected to be primordial. Scheila is likely to be a very old intact object and depleted in volatiles, at least at the surface level. This is consistent with our L-band observations, which show a featureless spectrum in the critical 3-$\mu$m region with no diagnostic absorption features found. The spectral differences between Scheila and other MBCs suggests that it may have a distinctly different composition from other known MBCs. We note that  these various physical differences between Scheila and other known MBCs do not rule out the possibility that its activity is cometary in nature.  They do however suggest that if Scheila's activity is cometary, the physical circumstances giving rise to that activity may be significantly different than those for other MBCs.

\subsection{Absence of Water ice and Possible Causes of the Activity}
Water ice is thermodynamically unstable on the surface of Scheila, where its sublimation rate of water ice greatly depends on its grain size and purity. Observations of 9P/Tempel 1 \citep{sunshine:2007} and new impact craters on Mars \citep{byrne:2009} consistently show that sub-surface ice exposed by impacts is relatively pure. The lifetime of a micron-sized pure ice grain at 3 AU is about 10$^{11}$s or 10$^3$ yr \citep{beer:2006}. However, if any impurities are present, this lifetime shortens drastically. Although our prompt NIR observations yielded no evidence of water ice or hydrated minerals, we can not conclude that Scheila contains no water ice on its surface. \cite{jewitt:2011} estimated that the upper limit of the area of exposed ice would be 100 km$^2$ if assuming an isothermal and spherical surface. A 100 km$^2$ ice patch occupies merely 0.25\% of the total surface area and would have been beyond the detectability of the observations reported here. 

If the sudden activity of asteroid (596) Scheila is powered by sublimation of water ice, we would then expect to see sublimating gas in the coma of this object. To investigate the possible cometary nature of Scheila, we made prompt measurements to search for the CN ($\Delta$v=0) band at 3880 \AA, which is the most easily detectable probe of volatile species in comets. We report these results in \cite{hsieh:2011b}. No evidence of the CN band in Scheila's spectrum was detected in these prompt Keck observations, corroborating results of UV-optical observations using the Swift telescope \citep{bodewits:2011}. The estimated upper limit in \cite{hsieh:2011b} for the CN production rate  and, in turn, the OH production rate ($Q_{OH}=2.93\times10^{26}$~s$^{-1}$), is consistent with the result ($Q_{OH}=2.04\times10^{26}$~s$^{-1}$) of \cite{howell:2011}. Although it is tempting to rule out the water ice sublimation process based solely on the non-detection of the CN emission and water ice absorption features, the sensitivity of the spectroscopic data is not sufficient to reach any solid conclusions. On the other hand, optical imaging of Scheila provides important clues about the origin of the comet-like activity. Using Hubble Space Telescope data, \cite{jewitt:2011} reported that the coma faded by about 30\% between two observations which were only 8 days apart. Such a short dissipation timescale is inconsistent with sublimation-driven dust emission, thus arguing for an impact excavation. Furthermore, the highly asymmetric structures and limb-brightening exhibited by Scheila's dust coma differs  from typical cometary comae, and can instead be better explained by a hollow cone that is produced via an impact \citep[{\it cf}.][]{hsieh:2011b}. Considering all the available observations, Scheila is best  described as a disrupted asteroid.  Our conclusion is consistent with the previous studies of Scheila \citep{jewitt:2011,bodewits:2011}. 

\section{Acknowledgment}
We thank David Jewitt and Norbert Schorghofer for their valuable discussions and constructive suggestions. BY was supported by the National Aeronautics and Space Administration through the NASA Astrobiology Institute under Cooperative Agreement No. NNA08DA77A issued through the Office of Space Science. HHH was supported by the National Aeronautics and Space Administration (NASA) through Hubble Fellowship grant HF-51274.01, awarded by the Space Telescope Science Institute, which is operated by the Association of Universities for Research in Astronomy, Inc., for NASA, under contract NAS 5-26555

\clearpage

\begin{deluxetable}{lllcccc}\tablewidth{4.0in}
\tabletypesize{\scriptsize}
\tablecaption{ Observational Parameters for (596) Scheila
  \label{obstable}}
\tablecolumns{6} \tablehead{  \colhead{UT Date }  & \colhead{r}  & \colhead{$\Delta$} &   \colhead{ $\alpha$}  &  \colhead{Total Exp.} & \colhead{Tel} &\colhead{$\lambda$ Converage}  \\
\colhead{}&\colhead{(AU)}&\colhead{(AU)}&\colhead{deg}&\colhead{s}&\colhead{}&\colhead{$\mu$m}}
\startdata
2010 Dec. 13 &  3.106 &  2.516 &16.2& 800 & Subaru & 1.4 - 4.0 \\
2010 Dec. 14 &  3.104 & 2.502 &16.1&1600 & Subaru & 1.4 - 4.0 \\
2011 Jan. 4 &  3.073 &  2.257 &12.0& 400 & IRTF & 0.8 - 2.5 \\
2011 Jan. 5 & 3.072 & 2.248 &11.7& 400 & IRTF & 0.8 - 2.5  \\
\enddata
\end{deluxetable}

\clearpage
\begin{thebibliography}{36}
\expandafter\ifx\csname natexlab\endcsname\relax\def\natexlab#1{#1}\fi

\bibitem[{Abell {et~al.}(2005)Abell, Fern{\'a}ndez, Pravec, French, Farnham,
  Gaffey, Hardersen, Ku{\v s}nir{\'a}k, {\v S}arounov{\'a}, Sheppard, \&
  Narayan}]{Abell:2005}
Abell, P.~A., {et~al.} 2005, Icarus, 179, 174

\bibitem[{A'Hearn(1988)}]{AHearn:1988}
A'Hearn, M.~F. 1988, Annual Review of Earth and Planetary Sciences, 16, 273

\bibitem[{Beck {et~al.}(2011)Beck, Quirico, Sevestre, Montes-Hernandez,
  Pommerol, \& Schmitt}]{Beck:2011}
Beck, P., Quirico, E., Sevestre, D., Montes-Hernandez, G., Pommerol, A., \&
  Schmitt, B. 2011,\aap, 526, 85

\bibitem[Beer et al.(2006)]{beer:2006} Beer, E.~H., Podolak, M., 
\& Prialnik, D.\ 2006, \icarus, 180, 473 

\bibitem[Bodewits et al.(2011)]{bodewits:2011} Bodewits, D., Kelley, 
M.~S., Li, J.-Y., Landsman, W.~B., Besse, S., 
\& A'Hearn, M.~F.\ 2011, \apjl, 733, L3 

\bibitem[{Bottke {et~al.}(2005)Bottke, Durda, Nesvorn{\'y}, Jedicke,
  Morbidelli, Vokrouhlick{\'y}, \& Levison}]{bottke:2005}
Bottke, W.~F., Durda, D.~D., Nesvorn{\'y}, D., Jedicke, R., Morbidelli, A.,
  Vokrouhlick{\'y}, D., \& Levison, H. 2005, Icarus, 175, 111

\bibitem[Byrne et al.(2009)]{byrne:2009} Byrne, S., et al.\ 2009, 
Science, 325, 1674 

\bibitem[{Bus \& Binzel(2002)}]{bus:2002}
Bus, S.~J., \& Binzel, R.~P. 2002, Icarus, 158, 146

\bibitem[{Campins {et~al.}(2010)Campins, Hargrove, Pinilla-Alonso, Howell,
  Kelley, Licandro, Moth{\'e}-Diniz, Fern{\'a}ndez, \& Ziffer}]{campins:2010}
Campins, H., {et~al.} 2010, Nature, 464, 1320

\bibitem[{Cushing {et~al.}(2004)Cushing, Vacca, \& Rayner}]{cushing:2004}
Cushing, M.~C., Vacca, W.~D., \& Rayner, J.~T. 2004, \pasp, 116, 362

\bibitem[{Dorschner {et~al.}(1995)Dorschner, Begemann, Henning, Jaeger, \&
  Mutschke}]{dorschner:1995}
Dorschner, J., Begemann, B., Henning, T., Jaeger, C., \& Mutschke, H. 1995,
 \aap, 300, 503

\bibitem[{Fern\`andez {et~al.}(2003)Fern\`andez, Sheppard, \&
  Jewitt}]{fernandez:2003}
Fern\`andez, Y.~R., Sheppard, S.~S., \& Jewitt, D.~C. 2003, \aj, 126, 1563

\bibitem[{Hapke(1981)}]{hapke:1981}
Hapke, B., 1981, \jgr, 86, 3039

\bibitem[{Hapke(1993)}]{hapke:1993}
Hapke,B., 1993, Theory of reflectance and emittance spectroscopy

\bibitem[{Henning {et~al.}(1999)Henning, Il'In, Krivova, Michel, \&
  Voshchinnikov}]{Henning:1999}
Henning, T., Il'In, V.~B., Krivova, N.~A., Michel, B., \& Voshchinnikov, N.~V.
  1999, Astronomy and Astrophysics Supplement Series, 136, 405

\bibitem[Hiroi et al.(2003)]{hiroi:2003} Hiroi, T., et al.\ 2003, 
Lunar and Planetary Institute Science Conference Abstracts, 34, 1425 

\bibitem[Howell 
\& Lovell(2011)]{howell:2011} Howell, E.~S., \& Lovell, A.~J.\ 2011, \iaucirc, 9191, 2 

\bibitem[Hsieh et al.(2004)]{hsieh:2004} Hsieh, H.~H., Jewitt, 
D.~C., \& Fern{\'a}ndez, Y.~R.\ 2004, \aj, 127, 2997 

\bibitem[{Hsieh \& Jewitt(2006)}]{hsieh:2006}
Hsieh, H.~H., \& Jewitt, D. 2006, Science, 312, 561

\bibitem[Hsieh et al.(2009a)]{hsieh:2009a} Hsieh, H.~H., Jewitt, D., 
\& Fern{\'a}ndez, Y.~R.\ 2009, \apjl, 694, L111 

\bibitem[{Hsieh {et~al.}(2009b)Hsieh, Jewitt, \& Ishiguro}]{hsieh:2009b}
Hsieh, H.~H., Jewitt, D., \& Ishiguro, M. 2009, \aj, 137, 157

\bibitem[Hsieh et al.(2010)]{hsieh:2010} Hsieh, H.~H., Jewitt, D., 
Lacerda, P., Lowry, S.~C., \& Snodgrass, C.\ 2010, \mnras, 403, 363 

\bibitem[Hsieh et al.(2011a)]{hsieh:2011a} Hsieh, H.~H., Meech, 
K.~J., \& Pittichova, J.\ 2011, arXiv:1106.0045 

\bibitem[Hsieh {\it et al.}(2011b)]{hsieh:2011b} Hsieh, H.~H., Yang, B., \& Haghighipour, N.
  2011, submitted to \apjl

\bibitem[Jewitt et al.(2011)]{jewitt:2011} Jewitt, D., Weaver, H., 
Mutchler, M., Larson, S., \& Agarwal, J.\ 2011, arXiv:1103.5456 

\bibitem[Jewitt et al.(2009)]{jewitt:2009} Jewitt, D., Yang, B., 
\& Haghighipour, N.\ 2009, \aj, 137, 4313 

\bibitem[{Jones {et~al.}(1990)Jones, Lebofsky, Lewis, \& Marley}]{jones:1990}
Jones, T.~D., Lebofsky, L.~A., Lewis, J.~S., \& Marley, M.~S. 1990, Icarus, 88,
  172

\bibitem[{Keller {et~al.}(2004)Keller, Britt, Buratti, \& Thomas}]{Keller:2004}
Keller, H.U., Britt, D., Buratti, B.J. and Thomas, N., 2004, in Comets II, ed. M.C. Festou, H.U. Keller and 
H.A. Weaver, (Tucson, AZ: Univ. Arizona Press), 222

\bibitem[Larson(2010)]{larson:2010} Larson, S.~M.\ 2010, \iaucirc, 
9188, 1

\bibitem[{Lebofsky(1980)}]{lebofsky:1980}
Lebofsky, L.~A. 1980, \aj, 85, 573

\bibitem[{Licandro {et~al.}(2003)Licandro, Campins, Hergenrother, \&
  Lara}]{Licandro:2003}
Licandro, J., Campins, H., Hergenrother, C., \& Lara, L.~M. 2003, Astronomy and
  Astrophysics, 398, L45

\bibitem[{Licandro {et~al.}(2011)Licandro, Campins, Kelley, Hargrove,
  Pinilla-Alonso, Cruikshank, Rivkin, \& Emery}]{licandro:2011}
Licandro, J., Campins, H., Kelley, M., Hargrove, K., Pinilla-Alonso, N.,
  Cruikshank, D., Rivkin, A.~S., \& Emery, J. 2011,\aap,
  525, 34

\bibitem[Mastrapa et al.(2009)]{mastrapa:2009} Mastrapa, R.~M., 
Sandford, S.~A., Roush, T.~L., Cruikshank, D.~P., 
\& Dalle Ore, C.~M.\ 2009, \apj, 701, 1347 

\bibitem[Minowa et al.(2010)]{minowa:2010} Minowa, Y., et al.\ 
2010, \procspie, 7736,  122

\bibitem[{Morrison(1973)}]{morrison:1973}
Morrison, D. 1973, Icarus, 19, 1

\bibitem[{Rayner {et~al.}(2003)Rayner, Toomey, Onaka, Denault, Stahlberger,
  Vacca, Cushing, \& Wang}]{rayner:2003}
Rayner, J.~T., Toomey, D.~W., Onaka, P.~M., Denault, A.~J., Stahlberger, W.~E.,
  Vacca, W.~D., Cushing, M.~C., \& Wang, S. 2003, \pasp, 115, 362

\bibitem[{Rivkin \& Emery(2010)}]{rivkin:2010}
Rivkin, A.~S., \& Emery, J.~P. 2010, Nature, 464, 1322

\bibitem[{Rivkin {et~al.}(2002)Rivkin, Howell, Vilas, \&
  Lebofsky}]{rivkin:2002}
Rivkin, A. S., Howell, E. S., Vilas, F., \& Lebofsky, L. A. 2002, in Asteroids III, ed. W. F. Bottke, A. Cellino Jr., P. Paolicchi, \& R. P. Binzel (Tucson, AZ: Univ. Arizona Press), 235

\bibitem[Prialnik 
\& Rosenberg(2009)]{prialnik:2009} Prialnik, D., \& Rosenberg, E.~D.\ 2009, \mnras, 399, L79 

\bibitem[{Rousselot {et~al.}(2011)Rousselot, Dumas, \& Merlin}]{rousselot:2011}
Rousselot, P., Dumas, C., \& Merlin, F. 2011, Icarus, 211, 553

\bibitem[Schorghofer(2008)]{schorghofer:2008} Schorghofer, N.\ 2008, 
\apj, 682, 697 

\bibitem[{Sunshine {et~al.}(2007)Sunshine, Groussin, Schultz, A'Hearn, Feaga,
  Farnham, \& Klaasen}]{sunshine:2007}
Sunshine, J.~M., Groussin, O., Schultz, P.~H., A'Hearn, M.~F., Feaga, L.~M.,
  Farnham, T.~L., \& Klaasen, K.~P. 2007, Icarus, 191, 73

\bibitem[{Tedesco {et~al.}(2002)Tedesco, Noah, Noah, \& Price}]{tedesco:2002}
Tedesco, E.~F., Noah, P.~V., Noah, M., \& Price, S.~D. 2002, \aj, 123, 1056

\bibitem[{Tholen \& Barucci(1989)}]{tholen:1989}
Tholen, D. J., \& Barucci, M. A. 1989, in Asteroids II, ed. R. Binzel et al. (Tucson: Univ. Arizona Press), 298 

\bibitem[{Warren(1984)}]{warren:1984}
Warren, S.~G. 1984, Applied Optics, 23, 1206

\bibitem[{Yang {et~al.}(2009)Yang, Jewitt, \& Bus}]{yang:2009}
Yang, B., Jewitt, D., \& Bus, S.~J. 2009, \aj, 137, 4538

\bibitem[Yang 
\& Sarid(2010)]{yang:2010} Yang, B., \& Sarid, G.\ 2010, \iaucirc, 9139, 2 

\bibitem[{Zellner {et~al.}(1985)Zellner, Tholen, \& Tedesco}]{zellner:1985}
Zellner, B., Tholen, D.~J., \& Tedesco, E.~F. 1985, Icarus, 61, 355

\end{thebibliography}
\end{document}